\documentclass[submission, Proceedings,11pt,A4paper]{SciPost}

\begin{document}

\begin{center}{\Large \textbf{
Clustering in $^{18}$O - absolute determination of branching ratios via high-resolution particle spectroscopy
}}\end{center}

\begin{center}
S. Pirrie\textsuperscript{1*},
C. Wheldon\textsuperscript{1},
Tz. Kokalova\textsuperscript{1},
J. Bishop\textsuperscript{1},
R. Hertenberger\textsuperscript{2},
H.-F. Wirth\textsuperscript{2},
S.~Bailey\textsuperscript{1},
N. Curtis\textsuperscript{1},
D. Dell'Aquila\textsuperscript{3},
Th. Faestermann\textsuperscript{4},
D. Mengoni\textsuperscript{5},
R. Smith\textsuperscript{1},
D.~Torresi\textsuperscript{1},
A. Turner\textsuperscript{1}
\end{center}

\begin{center}
{\bf 1} School of Physics and Astronomy, University of Birmingham
\\
{\bf 2} Fakult\"{a}t f\"{u}r Physik, Ludwig-Maximilians-Universit\"{a}t M\"{u}nchen
\\
{\bf 3} Universit\`a degli Studi di Napoli Fedorico II
\\
{\bf 4} Physik Department, Technische Universit\"{a}t M\"{u}nchen
\\
{\bf 5} Universit\`a degli Studi di Padova
\\
\vspace{0.5cm}
* Corresponding author email: S.Pirrie@PGR.bham.ac.uk
\end{center}

\begin{center}
\today
\end{center}


\section*{Abstract}
{\bf
The determination of absolute branching ratios for high-energy states in light nuclei is an important and useful tool for probing the underlying nuclear structure of individual resonances: for example, in establishing the tendency of an excited state towards $\alpha$-cluster structure. Difficulty arises in measuring these branching ratios due to similarities in available decay channels, such as ($\mathbf{^{18}}$O,$\mathbf{n}$) and ($\mathbf{^{18}}$O,$\mathbf{2n}$), as well as differences in geometric efficiencies due to population of bound excited levels in daughter nuclei. Methods are presented using Monte Carlo techniques to overcome these issues.
}

\section{Introduction}
\label{sec:intro}
The arrangement of nucleons in the nuclear system, especially in cases when these arrangements present physics outside of the largely successful nuclear shell model, is of great interest and importance in the elucidation of the nuclear force. Nuclear clustering, particularly $\alpha$-clustering, also provides a useful tool in the testing of ab-initio models, enabling many-nucleon systems to be modelled due to a reduction in required processing power. In recent years, there has been increased interest due to the potential existence of nuclear molecules outside of the $N=Z$ nuclei where $\alpha$-clustering has been well established \cite{Freer:2010}. Nuclear molecules are cluster structures in which increased stability is achieved between two clusters due to binding provided by valence nucleons, typically neutrons due to the lack of Coulomb repulsion. The $^{18}$O nucleus presents as a strongly viable candidate for $\alpha$-cluster structure, with many theoretical and experimental suggestions of potential cluster states \cite{vonOertzen2009,Johnson:2009kj,avila,artemov,MORGAN,curtis,fortune}. The likelihood of clustering in $^{18}$O is in part due to both the properties of the $^{12}$C and $^{14}$C nuclei: as $^{12}$C presents with a doubly-closed $p_{3/2}$ sub-shell, it has an increased stability accentuated by the energy of its first excited state, 4.4 MeV, which indicates the potential of $^{12}$C as a good cluster; the $^{14}$C nucleus is similar in this regard, with complete neutron and partial proton $p$-shell closure, and has a first excited state at above 6 MeV.

Experimental work performed by von Oerzten \textit{et al}.\ \cite{vonOertzen2009} found 30 new resonances in $^{18}$O through use of the $^{12}$C($^{7}$Li,$p$)$^{18}$O$^{*}$ reaction, measuring the energy of the recoil proton with the Q3D magnetic spectrograph at the Maier-Leibnitz Laboratory in Munich. From these and previously measured states, rotational fitting was performed from which several rotational bands were tentatively assigned. Of proposed these rotational bands, two have proposed cluster structure, each presenting as a parity doublet due to broken intrinsic reflection symmetry: the $K =  0_{2}^{+/-}$ with a $^{14}$C $\bigotimes$ $\alpha$ (core + $\alpha$) structure \cite{MORGAN,artemov} and the $K =  0_{4}^{+/-}$ with a $^{12}$C $\bigotimes$ 2$n$ $\bigotimes$ $\alpha$ (nuclear molecule) structure \cite{curtis,fortune}.

In order to determine the validity of these rotational bands, the tendency towards $\alpha$-clustering must be established by comparison of the reduced partial $\alpha$-width, $\gamma_{\alpha}^2$, to the Wigner limit. This limit is a theoretical reduced partial $\alpha$-width representing a case in which, for a particular excited state of a nucleus, an $\alpha$-particle is permanently preformed inside of the nucleus \cite{rmatrixbible}. The Wigner limit, $\gamma^2_{W}$, is defined by

\begin{equation}
    \gamma^2_{W} = \frac{3\hbar^2}{2\mu\alpha^2},
\end{equation}

\noindent where $\mu$ is the reduced mass of the two-body $\alpha$-decay exit channel and $\alpha$ is the channel radius. In order to determine $\gamma_{\alpha}^2$ for excited states in $^{18}$O, the absolute branching ratios for these high-energy states had to be measured. The absolute branching ratio could then be used to determine the partial $\alpha$-width, from which $\gamma_{\alpha}^2$ could be calculated by removing penetrability contributions, $P_l$, via the relationship

\begin{equation}
    \gamma_\alpha^2 = \frac{\Gamma_\alpha}{2P_l}.
\end{equation}

\noindent By taking the ratio $\gamma_\alpha^2/\gamma^2_{W}$, the tendency towards preformation of an $\alpha$-particle and hence towards $\alpha$-clustering can be established. For states in a cluster band, a ratio of larger than 0.1 would typically be expected, as well as a consistent value of this ratio across the composite states.

\section{Experimental method}

In order to measure the branching ratios of high-energy states in $^{18}$O, an experiment was performed at the Maier-Leibnitz Laboratory, again utilising the $^{12}$C($^{7}$Li,$p$)$^{18}$O$^{*}$ reaction in order to populate the relevant states. A 44 MeV $^{7}$Li provided by the Van de Graaff tandem accelerator at MLL was incident on a 110 $\mu$g/cm$^2$ natural carbon target, placed in the experimental chamber under vacuum. The Q3D magnetic spectrograph was set at an angle of $-$39$^{\circ}$ in-plane and 0$^{\circ}$ out-of-plane, with angular acceptances of $\pm$3.0$^{\circ}$ and $\pm$2.0$^{\circ}$ respectively, and a focal plane detector positioned after the quadrupole and dipole magnets. The focal plane detector consisted of two cathode foils, one segmented into 255 strips in order to give a position measurement for incident particles, as well as two anode wires and a scintillator detector. The anode wires in conjunction with the cathode foil were able to measure energy loss of a charged particle, while the scintillator measured the total deposited energy \cite{wirth}. This allowed for particle identification of incident protons, confirming that the correct reaction channel was observed. The position of the incident protons measured on the segmented cathode foil was proportional to its energy as it entered the Q3D spectrograph, and from this value the excitation functions of $^{18}$O were produced.

\begin{figure}[ht!]
\begin{minipage}{0.45\linewidth}
\includegraphics[width=\linewidth]{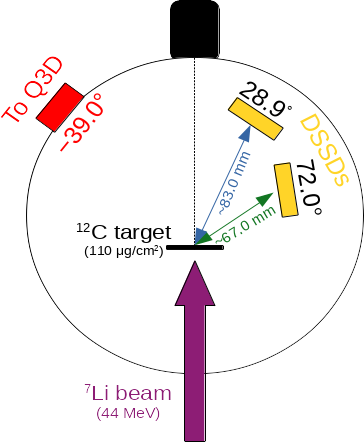}
\end{minipage}\hspace{3.5pc}%
\begin{minipage}{0.45\linewidth}
\includegraphics[width=\linewidth]{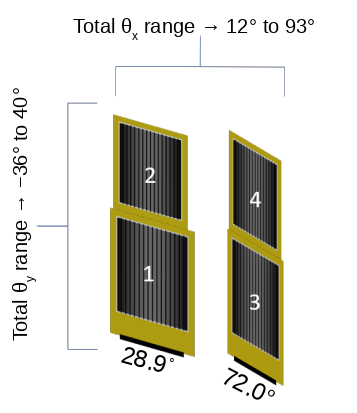}
\end{minipage} 
\caption{A diagram depicting the experimental set-up used in the chamber, showing the positions of the Q3D magnetic spectrograph and DSSD array (left). The arrangement of the constituent DSSDs in the Birmingham DSSD array, as well as the total angular coverage, is also shown (right) \cite{pirrie1,pirrie2}.}
\end{figure}

Once the correct reaction had been selected, the products from the decay of the $^{18}$O were detected through use of the Birmingham double-sided silicon strip (DSSD) array. This array consists of four 500 $\mu$m thick W1 Micron Semiconductor Ltd DSSDs \cite{micron}, with 16 horizontal and 16 vertical strips spanning a total active area of 50 mm x 50 mm, able to give both energy and position information on detected charged particles to enable high-resolution reconstruction of decay products. Two of the detectors were placed at an in-plane centre angle of 28.9$^{\circ}$ ($\approx$67.0 mm away from the target), while the other two were placed at a centre angle of 72.0$^{\circ}$ ($\approx$83.0 mm away from the target). The total angular coverage of the array was 12$^{\circ}\rightarrow$ 93$^{\circ}$ in-plane and $-$36$^{\circ}\rightarrow$ 40$^{\circ}$ out-of-plane, allowing for the detection of particles over a large solid angle and the detection of recoil $^{18}$O from the $\gamma$-decay of bound states.

\section{Reconstruction of decay products}

The excited states in $^{18}$O investigated lay between 7 MeV and 16 MeV, and as such there were several decay channels that had to be considered in the case of the higher-lying excitations. The $\alpha$-decay channel becomes available at an excitation energy of 6.228 MeV, so that $\gamma_{\alpha}^2$ of all states that were investigated was able to be determined. Comparatively, the other decay channels occur at the following excitation energies: $n$-decay at 8.045 MeV, 2$n$-decay at 12.19 MeV and $p$-decay at 15.94 MeV. In order to determine the species of particle detected by the DSSD array, a kinematic method was employed \cite{wheldon}. By using the measured energy and position of a particle, conservation of both energy and momentum could be employed in order to determine its species through use of a Catania plot. For the scenario in which the product $^{18}$O nucleus decays into particles $A^*$ and $B$, the $Q$-value expression for the total reaction can be rearranged to give

\begin{equation} \label{eq:q}
    E_{beam} - E_A - E_p = E_B + E_x - Q, 
\end{equation}
\noindent where $E_x$ is the excitation energy of particle $A^*$, $E_p$, $E_A$ and $E_B$ are the kinetic energies of the proton incident on the Q3D, $A^*$ and $B$ respectively. If $A^*$ is detected in the DSSD array, its momentum can be determined through the detection position and the measured kinetic energy, as long as the correct mass $m_A$ is assumed. Then, via conservation of momentum, the total momentum of $B$, $p_B$, can be determined and related to $E_B$ through $E_B = \frac{p_B^2}{2m_B}$. Equation \ref{eq:q} can then be adapted to give the following expression:

\begin{equation} \label{eq:cat}
    E_{beam} - E_A - E_p = \frac{1}{m_B} \frac{p_B^2}{2} - Q_3,
\end{equation}

\noindent $Q_3$ being the effective $Q$-value given by $Q - E_x$. By plotting the known quantity $E_{beam} - E_A - E_p$ against the calculated quantity $\frac{p_B^2}{2}$, a locus demonstrating a linear relationship is obtained with gradient $\frac{1}{m_B}$ and $y$-intercept $-Q_3$ if the assumption made about the mass, $m_A$, is correct. If this assumption is not correct, this locus will not lie on the appropriate line, but can be identified through comparison with Monte Carlo data. An example of this is shown later in Figure \ref{fig:cat}, in which it is assumed the $^{18}$O has decayed into $^{14}$C (detected) and an $\alpha$-particle.
\subsection{Kinematic cones}

As the initial reaction involves a two-body exit channel, the proton and the excited $^{18}$O nucleus, for a particular populated resonance the available energy is shared by the two final state particles in the same way every time. Events of interest, in which a charged particle is measured at the DSSD array in time coincidence with a proton measured at the focal plane detector, are hence constrained by the path of the proton into the angular acceptance of the Q3D. 

\begin{figure}[ht!]
\begin{minipage}{0.55\linewidth}
\includegraphics[width=\linewidth]{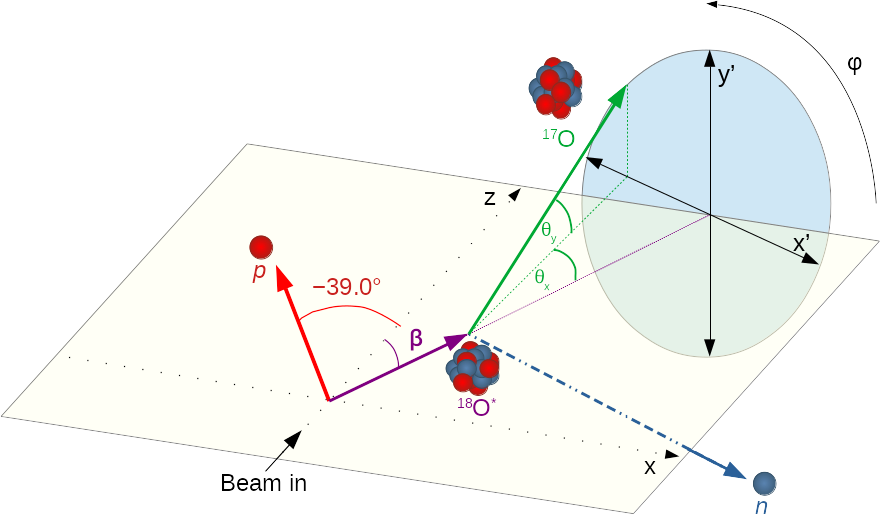}
\end{minipage}\hspace{1.8pc}%
\begin{minipage}{0.4\linewidth}
\includegraphics[width=\linewidth]{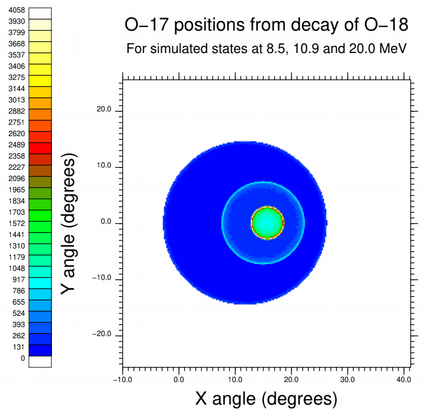}
\end{minipage} 
\caption{A diagram depicting the total reaction $^{12}$C($^{7}$Li,$p$)$^{17}$O$+n$, showing the area in which the $^{17}$O is constrained (a cross-sectional slice of the kinematic cone) due to detection of the proton within the angular acceptance of the Q3D (left). Also shown are Monte Carlo data, generated for three different $^{18}$O excitation energies, showing the increase in the radius of these kinematic cones for an increase in excitation energy (right).}
\label{fig:kin}
\end{figure}

Figure \ref{fig:kin} shows a depiction of an event, with the beam direction along the $z$-axis and target position at the origin, in which the product $^{18}$O$^*$ nucleus emits a neutron to form $^{17}$O in the ground state and the proton ejectile enters in the centre of the angular acceptance of the Q3D ($-$39.0$^{\circ}$ in-plane and 0$^{\circ}$ out-of-plane). As a result, the $^{18}$O$^*$ is constrained to the $x$-$z$ plane to conserve momentum out-of-plane, and travels at an angle $\beta$ in-plane depending on the amount of kinetic energy available: for a larger excitation energy in $^{18}$O$^*$, a smaller $\beta$ would be observed. In the next step, the two-body decay is constrained to another two-dimensional plane, not necessarily the same as the $x$-$z$ plane. The blue circle in Figure \ref{fig:kin} shows the kinematically allowed region in which the $^{17}$O could be detected in a plane perpendicular to the path of the $^{18}$O$^*$. In fact, the path of the $^{17}$O is thus constrained within the volume of a cone which has a radius dependent on the excitation energy of the $^{18}$O; the most likely position is however on the surface area of the cone due to kinematic focusing effects.

Also in Figure \ref{fig:kin} is an example of a Monte Carlo simulation, produced using the in-house nuclear reaction kinematics code \textsc{Resolution8.1} (see references \cite{pirrie2} \cite{res} \cite{res2} for more detail), of the described interaction for different excitation energies. This Monte Carlo code simulates nuclear reactions as a series of sequential two-body decays, allowing for a variety of real world smearing effects to also be simulated in order to best reproduce experimental data. The kinematic focusing effects, as well as the increase in the radius of the kinematic cones and the decrease in angle $\beta$ for greater excitation energies, can be clearly seen. The difference in the radii of these cones gives the ability for distinction between same-particle decay paths through different states from the hit pattern on the detector arrays.

\subsection{Excited levels in daughter nuclei}

More complicated decay paths than the one depicted in Figure \ref{fig:kin} are also possible: for example, decays of $^{18}$O that populate excited levels in the daughter nuclei. If the daughter nucleus is excited, the available kinetic energy for the two decay fragments is correspondingly decreased and thus a kinematic cone with a smaller radius is observed. Due to the importance of accurately determining the expected geometric efficiency of the DSSD array for the detection of decay products, it is thus paramount to determine whether or not these levels are populated. 

\vspace{5mm}

\begin{figure}[ht!]
\begin{minipage}[ht!]{0.495\linewidth}
  \centering
  \includegraphics[width=17pc]{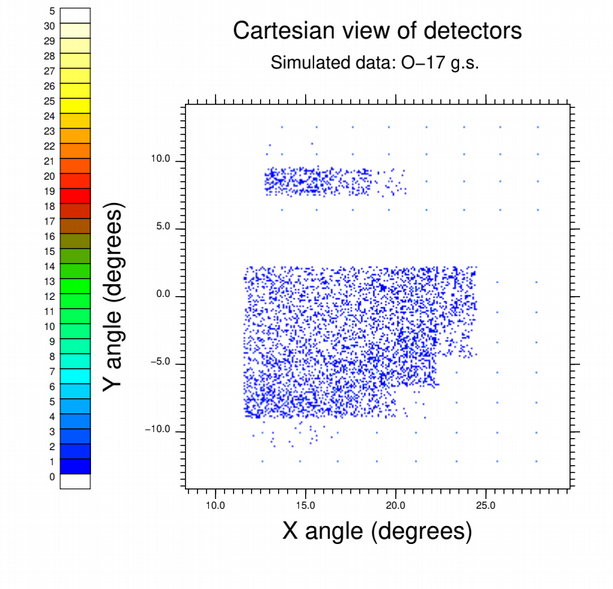}
\end{minipage}
\begin{minipage}[ht!]{0.495\linewidth}
  \centering
  \includegraphics[width=17pc]{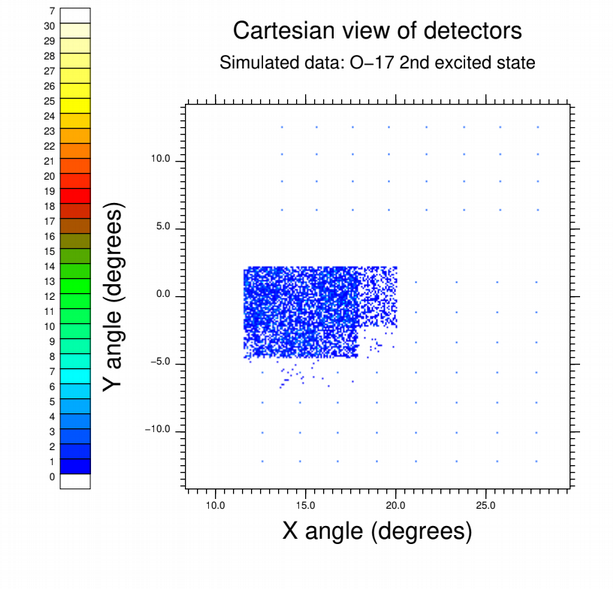}
\end{minipage}
\begin{minipage}{\linewidth}
  \centering
  \includegraphics[width=17pc]{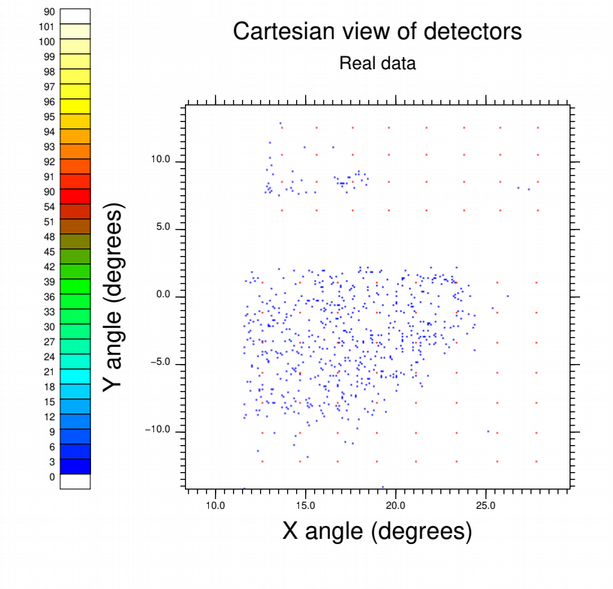}
\end{minipage}
  \caption{A comparison of Monte Carlo data for the $n$-decay of the 11.855 MeV state in $^{18}$O, both to the ground state (top left) and the 3.055 MeV state (top right) of $^{17}$O, to real data (bottom). A much better agreement to the ground state Monte Carlo data can be observed. The centres of the individual pseudo-pixels are also indicated by the grid of dots.}
    \label{fig:xy}
\end{figure}

\subsubsection{Determination through hit position on detectors} \label{sec:xy}

It is possible, through the use of position sensitive charged-particle detectors such as the DSSD array, to distinguish between cases of different excitation energy in the daughter nucleus by looking at the hit positions of any charged decay fragments. An example of this is shown in Figure \ref{fig:xy}, for the case of decay of an $^{18}$O nucleus with an excitation energy of 11.855 MeV which then subsequently $n$-decays into $^{17}$O. Monte Carlo data are presented for both the population of the ground state and for the 2\textsuperscript{nd} excited state at 3.055 MeV and compared with the hit positions for real data. The latter were gated on both the 11.855 MeV state as measured by the focal plane detector and on $n$-decay events as determined via a Catania plot. It is clear that the majority of decays for this case populate the ground state of $^{17}$O. The ability to distinguish between the different excitation energies relies mostly the solid angle subtended by pixels, as for very small differences in excitation energy one would rely on the circumference of the kinematic cone to move across a pixel boundary. With a high level of statistics, it becomes easier to determine percentages for the population of these individual daughter states.

This method is sensitive to the excitation energy in $^{18}$O, as the radius of the kinematic cones change as the available kinetic energy for final state particles change. As such, it is important to look as these states individually if possible -- in the current work, this is achieved by choosing gates on the excitation function, produced by the recoil proton measured at the focal plane detector, of $^{18}$O that ideally include only events within 2$\sigma$ of the centroid measured for the relevant excited level. In rare cases where this cannot be achieved, two nearby states can be investigated together due to the small change in excitation energy resulting in a small change in kinematic distributions (and thus geometric efficiency) for each decay channel.

\subsubsection{Determination through Catania plots}

\begin{figure}[ht!]
\begin{minipage}{0.48\linewidth}
\includegraphics[width=\linewidth]{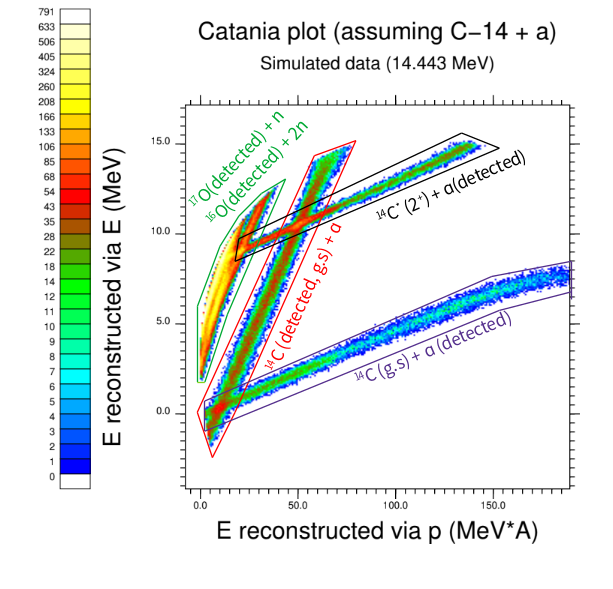}
\end{minipage}\hspace{1pc}%
\begin{minipage}{0.48\linewidth}
\includegraphics[width=\linewidth]{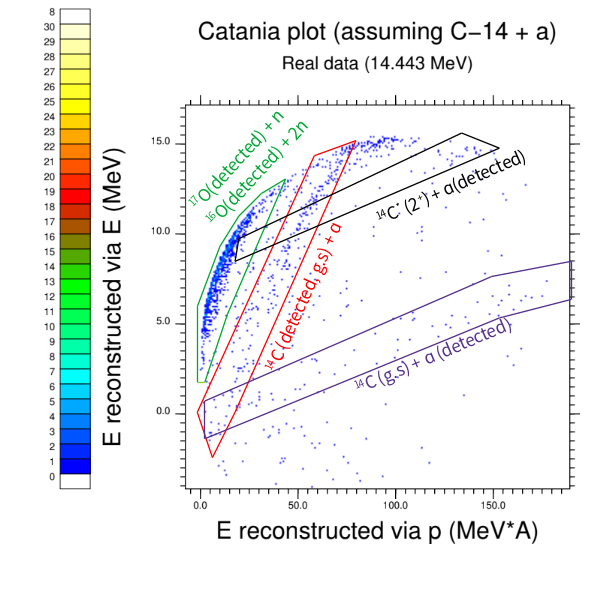}
\end{minipage} 
\caption{\label{fig:cat} A comparison of the Catania plots generated for $^{18}$O excited to 14.443 MeV, assuming an $\alpha$-decay and detected $^{14}$C, for Monte Carlo data (left) and real data (right). The individual decay channels are labelled in order to see the overlap of the loci.}
\end{figure}

It is also possible to determine the population of excited levels in daughter nuclei via Catania plots, as the $y$-intercept for loci associated with a particular decay channel depends on the excitation energy, as shown in Equation \ref{eq:cat}. An example of this is shown in Figure \ref{fig:cat}, in which both $\alpha$-decays to the ground state and 1\textsuperscript{st} excited state of $^{14}$C are presented alongside the other available decay channels, $n$-decay and $2n$-decay. The Catania plots in Figure \ref{fig:cat} are generated assuming a detected $^{14}$C, and the different loci are labelled. Despite the locus formed from events that populate the 1\textsuperscript{st} excited state lying across other loci, there are regions in which there is no contamination from other decay channels, enabling the determination of these branching ratios: by using Monte Carlo to establish the fraction of these events present in a uncontaminated region, the total number can be determined in the real data and hence appropriately subtracted from the other loci, giving more accurate values of the branching ratios for all channels.

This method, in conjunction with the method detailed in Section \ref{sec:xy}, can be used to reduce the uncertainty on the values for the individual branching ratios obtained. As the Catania plot can have regions in which there is no overlap between different loci, percentage population of different daughter levels can be obtained also.

\subsection{2n-decay}

For states in $^{18}$O above 12.19 MeV, the 2$n$-decay channel becomes energetically possible. This decay channel is harder to distinguish from others due to having two final state neutrons that are unable to be measured in the DSSD array. Events were simulated through Monte Carlo as sequential decay events passing through excited states above the $n$-decay threshold in $^{17}$O. 

\begin{figure}[ht!]
\begin{minipage}{0.48\linewidth}
\includegraphics[width=\linewidth]{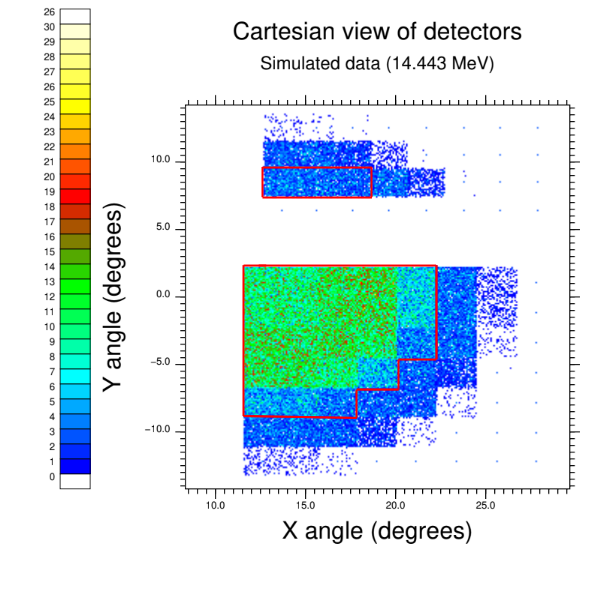}
\end{minipage}\hspace{1pc}%
\begin{minipage}{0.48\linewidth}
\includegraphics[width=\linewidth]{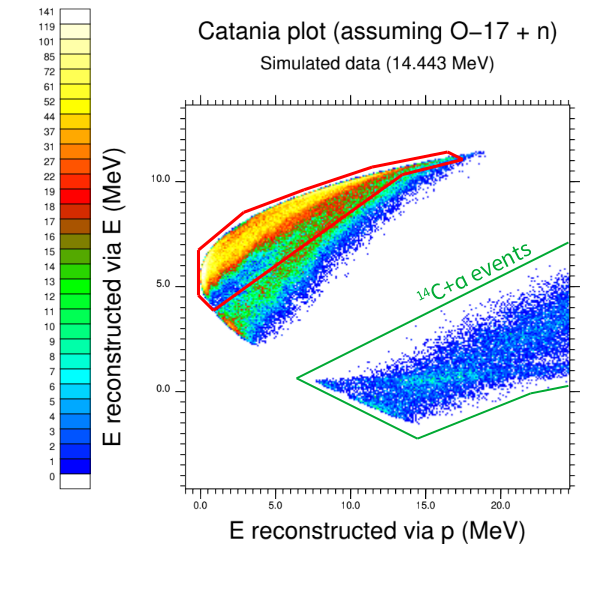}
\end{minipage} 
\caption{\label{fig:2n} A view of the hit positions of charged particles on the DSSD array (left) and a Catania plot generated assuming measured products arise from a $n$-decay of $^{18}$O$^*$ (right), both generated using Monte Carlo data for ($^{18}$O$^*$,$n$) and ($^{18}$O$^*$,2$n$). The regions highlighted in red represent areas within which $>$99.5\% of ($^{18}$O$^*$,2$n$) events are located.}
\end{figure}

In order to separate 2$n$-decay events from $n$-decay events, important again due to differing geometric efficiencies, use of both Catania plots and the hit positions of particles on the DSSD array can be used. However, due to the similarity of the two decay modes, their respective features have large overlaps. Using Monte Carlo to establish the fraction of $n$-decay events that lie within the regions defined by 2$n$-events, $f_n$, the total number of events from the $n$-decay channel, $N_n$, can be determined in the real data by measuring the number of $n$-decay events outside of these gates, $N_{out}$, and using the relationship

\begin{equation}
    N_n = \frac{N_{out}}{1-f_n}.
\end{equation}

\noindent Then, the actual number of events from the 2$n$-decay channel, $N_{2n}$ can be calculated by taking the total number of events inside of the gates, $N_{in}$, and subtracting the number of appropriate number of $n$-decay events through the expression

\begin{equation}
    N_{2n} = N_{in} - N_n \times f_n.
\end{equation}

\noindent An example of using a Catania plot and hit positions on the DSSD array for this scenario is shown in Figure \ref{fig:2n}, where the gates for 2$n$ events have been shown overlain on a mixture of single and double $n$-decay events. Using both in conjunction reduces the number of $n$-decay events that lie within the 2$n$-decay gate, reducing the uncertainty on the measured value $N_{out}$ - in the case of the 14.443 MeV state, $f_n$ is 39.0\%.

\begin{figure}[ht!]
\begin{minipage}{0.48\linewidth}
\includegraphics[width=\linewidth]{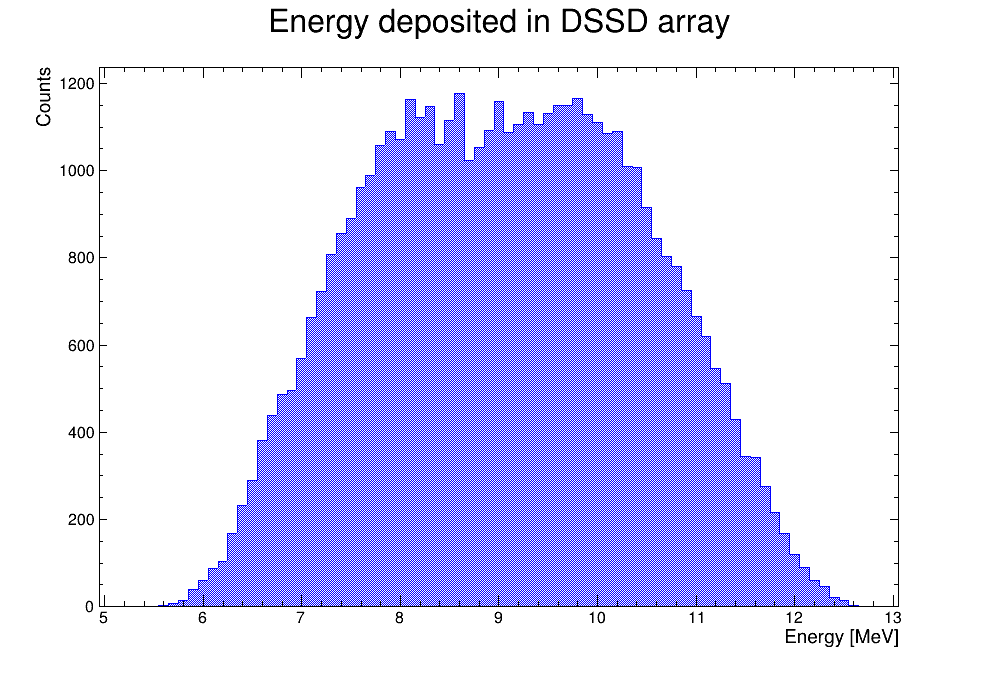}
\end{minipage}\hspace{1pc}%
\begin{minipage}{0.48\linewidth}
\includegraphics[width=\linewidth]{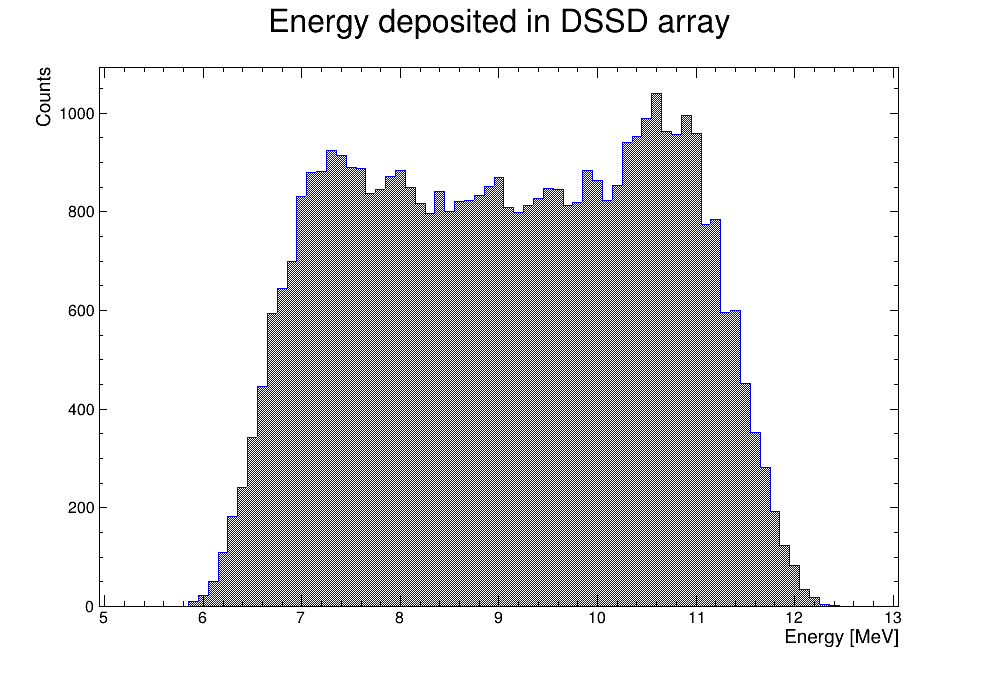}
\end{minipage} 
\caption{\label{fig:nn} A comparison of the distribution of energy deposited in the DSSD array by a product $^{16}$O through the $n$-decay of $^{17}$O$^*$ from the 4.553 MeV state (left) and the  6.356 MeV state (right).}
\end{figure}

Distinguishing between population of states that can $n$-decay in $^{17}$O is a more difficult task, as the resolution of the reconstruction of the final state particles decreases, due to a lack of detected particles in the intermediate $^{17}$O$^*$+$n$ stage. It is hence experimentally difficult to distinguish between levels populated above the $n$-threshold in $^{17}$O, especially with the statistics available in the current work. One possible method for distinguishing between these states that can be employed with higher statistics is presented in Figure \ref{fig:nn}, in which the energy of the measured $^{16}$O varies between different excitation energies in $^{17}$O$^*$. Information on $J^{\pi}$ of the parent state can also hint at the likelihood of populating various states above the $n$-decay threshold in $^{17}$O, as it is preferential for the $n$ to carry low units of angular momentum due to the decreased centrifugal barrier. 

\section{Conclusion}
A method for measuring high-energy branching ratios for states in $^{18}$O has been discussed, using the Q3D magnetic spectrograph in conjunction with the Birmingham DSSD array to allow high-resolution reconstruction of decay products, along with how these can be used to determine the tendency of states to $\alpha$-cluster. 

Also discussed was the importance of establishing whether various excited states in daughter nuclei are populated, due to a large change in geometric efficiency from the narrowed kinematic cone for the two-body decay products. Methods, both using hit positions arising as a direct result of these kinematic cones as well as Catania plots, a kinematic method for determining masses of detected particles without traditional particle identification from a DSSD telescope array, were presented for establishing the tendency towards population of these states.

Measuring sequential neutron decays with charged particle detectors and separating these events from other loci, again necessary due to the differing geometric efficiency of the two channels, has been discussed. Finally, a potential method for determining which states above the $n$-threshold are populated in a high statistics experiment was described.

\section*{Acknowledgements}
The authors would like to thank Andy Bergmaier for his help loaning equipment during the set-up of the experiment, as well as the operators of the tandem Van de Graaff accelerator at the Maier-Leibnitz Laboratory in Munich for providing and maintaining a stable $^{7}$Li beam.

\paragraph{Funding information}
This work was funded by the UK Science and Technology Facilities Council (STFC) under Grant No. ST/L005751/1 and from the European Union’s Horizon 2020 research and innovation programme under the Marie Sk\l{}odowska-Curie Grant Agreement No. 65F9744.

\bibliography{references.bib}

\nolinenumbers

\end{document}